\documentclass{aa}
\input epsf
\input psfig
\usepackage{graphics}
\usepackage{amssymb}
\usepackage{amsmath}

\newcommand{\apj}{ApJ}

\newcommand{\aap}{A\&A}
\newcommand{\araa}{ARA\&A}
\newcommand{\etal} {et al.}
\newcommand{\be}{\begin{equation}}
\newcommand{\ee}{\end{equation}}
\newcommand{\ba}{\begin{eqnarray}}
\newcommand{\ea}{\end{eqnarray}}
\begin{document}
\title{Submillimeter lines from circumstellar disks around pre-main
sequence stars}
\author{Gerd-Jan van Zadelhoff
\inst{1} \and Ewine F. van Dishoeck \inst{1} \and Wing-Fai Thi
\inst{1} \and Geoffrey A. Blake\inst{2}} 

\offprints{G.J. van Zadelhoff, zadelhof@strw.leidenuniv.nl} 

\institute{Leiden Observatory, P.O. Box 9513, 2300 RA
Leiden, The Netherlands, \and Division of Geological and Planetary
Sciences, California Institute of Technology, MS 150-21, Pasadena, CA
91125}

\abstract{ Observations of submillimeter lines of CO, HCO$^+$, HCN and
their isotopes from circumstellar disks around low mass pre-main
sequence stars are presented.  CO lines up to $J$=6$\to$5, and HCO$^+$
and HCN lines up to $J$=4$\to$3, are detected from the disks around
LkCa~15 and TW~Hya.  These lines originate from levels with higher
excitation temperatures and critical densities than studied before.
Combined with interferometer data on lower excitation lines, the line
ratios can be used to constrain the physical structure of the disk.
The different line ratios and optical depths indicate that most of the
observed line emission arises from an intermediate disk layer with
high densities of 10$^{6}-10^{8}$ cm$^{-3}$ and moderately warm
temperatures in the outer regions.  The data are compared with three
different disk models from the literature using a full 2D Monte Carlo
radiative transfer code. The abundances of the molecules are
constrained from the more optically thin $^{13}$C species and indicate
depletions of $\approx 1-30$ for LkCa~15 and very high depletions of
$> 100$ for TW~Hya with respect to dark cloud abundances.  Evidence
for significant freeze-out (factors of 10 or larger) of CO and HCO$^+$
onto grain surfaces at temperatures below 22~K is found, but the
abundances of these molecules must also be low in the warmer upper
layer, most likely as a result of photodissociation.  A warm upper
layer near the surface of a flaring disk heated by stellar and
interstellar radiation is an appropriate description of the
observations of TW~Hya. LkCa~15 seems to be cooler at the surface,
perhaps due to dust settling.  The density constraints are also well
fitted by the flared disk models.  } \authorrunning{van
Zadelhoff \etal} \titlerunning{Submillimeter lines from circumstellar
disks} \date{\today}

\maketitle

\section{Introduction}
Circumstellar disks play an essential role in the understanding of the
formation of planetary systems such as our own (see Beckwith 1999 and
Beckwith \& Sargent 1996 for recent reviews). These protoplanetary
disks contain a few percent of the mass of the pre-main sequence stars
which they surround.  One of the key questions concerning
circumstellar disks is their evolution toward planetary
formation. The different evolution scenarios can be constrained by
placing limits on the density and temperature distributions in the
disks.
The standard method for determining the disk physical structure
utilizes fits to the observed spectral energy distributions (SEDs)
(\cite{1988ApJ...326..865A}). This procedure relies on the changing
opacity of the dust at different wavelengths. At long wavelengths
(typically $\lambda >$1 mm), the dust emission is optically thin and
hence traces the product of mass and mean temperature (Beckwith 1999),
whereas at shorter wavelengths the disk becomes optically thick so
that only the temperature structure and geometry of the disk
surface-layer is probed.  The derived temperature and density solution
is not unique since different distributions or different dust
properties are able to fit the SEDs (\cite{1992A&A...263..113B},
\cite{1994A&A...287..493T}). In addition, changing dust properties
with position in the disk can affect the analysis, as can the disk
size.  Nevertheless, one of the more robust results has been the
recognition of relatively high temperatures in the surface layers of
the disk, implying that they need to be heated more efficiently by
stellar radiation compared to the traditional thin (flat) disk model.
This led \cite{1987ApJ...323..714K} to propose a flared disk geometry
in which the outer disk intercepts more radiation than does a flat
disk.

{\it Hubble Space Telescope} (HST) observations of young low mass
stars such as HH~30 and HK~Tauri show edge on (silhouette) disks which
indeed flare noticeably (\cite{1996ApJ...473..437B}).  The radiation
from the central star incident on the outer parts of the disk changes
the temperature and chemistry in those regions, with the temperature
change giving rise to a larger scale height and thereby flaring the
disk.  Recent models by Chiang \& Goldreich (1997, 1999) and D'Alessio
et al.\ (1997, 1998, 1999) include the irradiation of flared disks to
derive self-consistent models with a warm outer layer. The models by
Bell et al.\ (1997, 1999) take both the stellar radiation and
re-processing of radiation in the disk into account. The latter models
have an isothermal temperature in the vertical $z-$direction due to
large flaring in the inner disk region, thereby shielding the 
the outer disk from stellar
light.  Comparison with the other models provides a good test case
whether a high temperature upper layer is needed to satisfy the
observational constraints. All three types of models are used in this
work and will be discussed in more detail in \S \ref{sec: model}.

An alternative method to derive the density and temperature structure
in disks is through modeling of molecular line emission. Although the
inferred solution from observations of a single line is not unique,
data on a sufficiently large number of transitions of various
molecules can be used to constrain the temperature and density
independently. Moreover, careful analysis of the line profiles can
provide positional information, since the center of a line probes a
different radial part of the disk compared with the wings, unless the
disk is nearly face-on. In addition, observations of various
isotopomers can give information on different vertical regions of the
disks due to their varying optical depths. To date, most data concern
the lowest rotational $J$=1--0 and 2--1 transitions of $^{12}$CO and
$^{13}$CO, which originate from levels at low energies ($<20$~K) and
which have low critical densities ($< 5000$ cm$^{-3}$) (e.g., Dutrey
et al.\ 1996).  Data on molecules with larger dipole moments such as
HCN and HCO$^+$ have been limited to the 1.3 millimeter band (Dutrey
et al.\ 1997), except for the case of TW~Hya (Kastner et al.\ 1997).
In this paper, higher rotational lines in the 0.8 and 0.45 millimeter
atmospheric windows are presented, obtained with the {\it James Clerk
Maxwell Telescope} (JCMT) and {\it Caltech Submillimeter Observatory}
(CSO).  These lines probe higher temperatures (up to 100~K) and higher
densities (up to $10^7$ cm$^{-3}$) than do presently available
spectra.

The observations are accompanied by a detailed analysis of the
excitation and radiative transfer of the lines. In contrast with
previous models (e.g. G\'omez \& D'Alessio 2000), our analysis uses
statistical equilibrium (SE) rather than local thermodynamic
equilibrium (LTE) since the surface layers of the disk may have
densities below the critical density of various transitions.  
 In addition, the two-dimensional (2D) code
developed by Hogerheijde \& van der Tak (2000) is used to calculate
the full radiative transfer in the lines. The data
can be used to test the disk models described above that are fit to
the SEDs available for most T-Tauri and Herbig Ae stars.  In addition
to constraining the temperature and density, the observations and
models also provide information on the depletion of different species.

The molecular abundances and excitation are studied by comparing
different isotopomers of CO, HCO$^{+}$ and HCN for two sources: TW~Hya
and LkCa~15. TW~Hya is nearby (57 pc, Kastner et al. 1997) and has a
disk seen nearly face-on.  LkCa~15 is located at the edge of the
Taurus cloud at $\sim$140 pc and has an inclination of $\sim$60\degr,
where 0\degr \ is face-on.  Both sources show a wealth of molecular
lines and are well-suited for developing the analysis tools needed to
investigate disk structure.

The outline of this paper is as follows. In \S 2, we present the
observational data. In \S 3, we perform a simple zeroth-order
analysis of the observed line ratios to constrain the excitation
parameters. The adopted disk models are introduced in \S 4.1, whereas
the methods for calculating the level populations are explained in
\S 4.2--4.5. Finally, the results of the analysis are given in \S
5 and summarized in \S 6.

\section{Observations}
\label{sec obser1}
\begin{figure*}
\resizebox{\hsize}{!}{\includegraphics{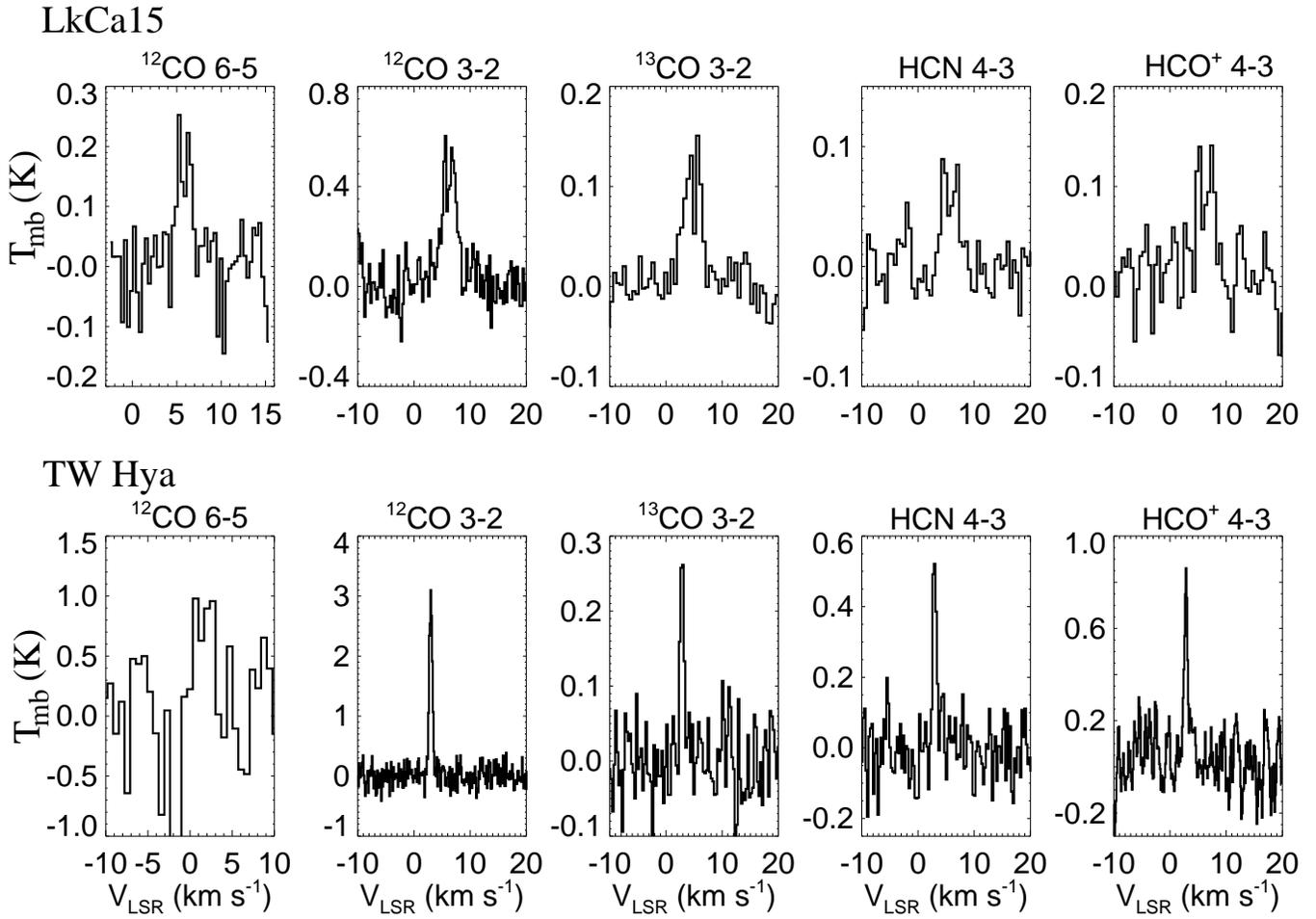}}
    \caption{\footnotesize Top: selected CO, HCO$^+$ and HCN lines
    toward LkCa~15. The profiles show a double-peaked structure
    typical for a disk seen at an inclination of about
    60$^{o}$. Bottom: selected CO, HCO$^+$ and HCN lines from the face-on disk
    around TW~Hya.}
\label{fig: obslkca}
\end{figure*}
Between September 1998 and December 2000, spectral line observations
were obtained for several pre-main sequence low mass stars surrounded by
circumstellar disks using the dual polarization B3 receiver at the
{\it James Clerk Maxwell Telescope} (JCMT\footnote{\footnotesize The
James Clerk Maxwell Telescope is operated by the Joint Astronomy
Centre in Hilo, Hawaii on behalf of the Particle Physics and Astronomy
Research Council in the United Kingdom, the National Research Council
of Canada and The Netherlands Organization for Scientific Research.})
in the 345 GHz (0.8 mm) band.  The observations were obtained mostly
in single side band (SSB) mode using beam-switching with a typical
switch of 180\arcsec \ in azimuth. The spectra were recorded with the
Digital Autocorrelation Spectrometer (DAS) at a frequency resolution
of $\sim$0.15 MHz ($\sim 0.15$ km s$^{-1}$), and were converted to the
main-beam temperature scale using $\eta_{\rm mb}$=0.62. See
{\tt $<$http://www.jach.hawaii.edu/JACpublic/JCMT/$>$} for details. The
calibration was checked regularly at each frequency setting against
standard spectra of bright sources obtained by the JCMT staff, and
were generally found to agree within 10\%.  Integration times ranged
from 30 minutes (ON+OFF) for $^{12}$CO 3--2 up to 120 minutes for
C$^{18}$O 3--2, reaching rms noise levels on the $T_{\rm mb}$ scale of
about 20 mK after adding the data from the two mixers together and
smoothing to 0.3 MHz resolution. A deep integration on the C$^{18}$O
2--1 line was obtained with receiver A3 for LkCa~15, which has
$\eta_{\rm mb}=0.79$.

These data are complemented by observations using the {\it Caltech
Submillimeter Observatory} (CSO\footnote{\footnotesize The Caltech
Submillimeter Observatory is operated by Caltech under a contract from
the National Science Foundation (NSF) AST-9981546}) of the $^{12}$CO
6--5 line for the same sources. In addition, the $^{12}$CO 2--1 line
has been observed with the IRAM 30m telescope\footnote{\footnotesize
Institute Radio Astronomie Millim\'etrique} for LkCa 15.  For the
$^{12}$CO 6--5 line, $\eta_{\rm mb}$=0.40, whereas for the IRAM
$^{12}$CO 2--1 observations the raw data are divided by 0.55 (see {\tt
$<$http://www.iram.es/$>$}).

Interferometer maps of the lowest rotational transitions of several
species toward LkCa~15 have been obtained by \cite{Qi} using the {\it
Owens Valley Millimeter Array} (OVRO). Some lines have also
been imaged by Simon et al.\ (2000) and Duvert et al.\ (2000)
with the IRAM Plateau de Bure
interferometer. In addition, the {\it Infrared
Space Observatory} (ISO) has detected the lowest rotational S(0) and
S(1) lines of H$_{2}$, which provide independent constraints on the
temperature and mass of warm gas and which are discussed elsewhere
(\cite{wf1}).  In this paper only the single dish results on CO,
HCO$^+$ and HCN for the sources LkCa~15 and TW~Hya
are presented.

Figure \ref{fig: obslkca} shows some of the spectra observed toward
LkCa~15 and TW~Hya. The double peaked profiles for LkCa~15 are
consistent with Keplerian rotation of the disk seen at an inclination
of 58 $\pm$ 10$^{\circ}$ (Qi 2000, Duvert et al.\ 2000). Since TW~Hya
is seen face-on, only narrow single-peaked lines are observed from
this source.  For both stars, the $^{12}$CO lines disappear  at
one beam offset from the source.  Table~1 summarizes the measured line
parameters and beam sizes at the observed frequencies.  The
upper-limits for LkCa~15 refer to a 2$\times$rms noise level, with the
limit on the integrated line strengths obtained by using two separate
gaussians each with a line-width of 1.3 km s$^{-1}$, as found for
$^{13}$CO 3-2.   For TW Hya, the upper limits assume a gaussian
with a width of 0.76 km s$^{-1}$, similar to that observed for HCN and
H$^{13}$CO$^{+}$ 4--3.  Note that our HCO$^+$ 4--3 line toward TW~Hya
is a factor of three weaker than that found by Kastner et al.\
(1997). We adopt our values in the analysis. The HCN 4--3 integrated
intensity is comparable to that found by Kastner et al.\ (1997) within
10 \%. There is a hint of a $^{12}$CO 6--5 line toward TW~Hya, but
this is treated as an upper limit.

\begin{center}
\begin{table*}
\caption{Observed line parameters for LkCa~15 and TW~Hya}
\begin{tabular}{lllllll}
\hline
line & $\int$ T$_{\rm mb}$ dv &T$_{\rm mb}$&  $\Delta V^{a}$ & Beam & Telescope & Date\\
     & K km s$^{-1}$      &K  &km s$^{-1}$&$\arcsec$ &  &  \\
\hline
\multicolumn{6}{c}{ LkCa~15} \\
\hline
CO 6--5        & 0.53      &0.29/0.28& 2.0   & 14.5         &CSO & Jun '00\\
CO 3--2        & 1.39      &0.60/0.56& 3.3 & 13.8           &JCMT& Nov '99\\
CO 3--2        & 1.17      &0.37/0.39& 2.2 & 25.7           &CSO & Feb '98\\
CO 2--1        & 1.82      &0.74/0.76& 2.9 & 10.5           &IRAM& Dec '98\\
$^{13}$CO 3--2 & 0.39      &0.13/0.15& 3.4 & 14.4           &JCMT& Sept '98\\
$^{13}$CO 1--0$^{e}$ & 7.43      &   &           &3.1$\times$2.6  &OVRO& \\
C$^{18}$O 3--2 &$<$0.14$^b$&$<$0.05$^{c}$& & 14.5         &JCMT& Nov '99 \\
C$^{18}$O 2--1 &$<$0.20$^b$&$<$0.07$^{c}$& & 22.2         &JCMT& Nov '98 \\
C$^{18}$O 1--0$^{e}$ & 0.70       &               & & 4.3$\times$4.0 &OVRO &  \\
HCO$^{+}$ 4--3 & 0.26      &0.14/0.14& 3.3 & 13.4           &JCMT& Sept '98\\
HCO$^{+}$ 1--0$^{e}$ & 4.19      &         &     & 4.5$\times$3.3 &OVRO&  \\
H$^{13}$CO$^{+}$ 4--3      &$<$0.13$^b$&$<$0.05$^{c}$& &13.7&JCMT& Jan '00 \\
H$^{13}$CO$^{+}$ 1--0$^{e}$& 0.07  &    & &13.0$\times$10.8  &OVRO&\\
HCN 4--3       & 0.25  &0.09/0.08& 3.3 &    13.5  &JCMT& Sept '98\\
HCN 1$_{2}-0_{1}^{e}$      & 3.04    &    & &4.3$\times$3.4  &OVRO& \\
H$^{13}$CN 3--2$^{e}$  & 1.49      &    & &0.9$\times$0.6  &OVRO& \\
H$^{13}$CN 1$_{2}-0_{1}^{e}$& 1.20    &    & &5.8$\times$4.6  &OVRO& \\
\hline
\multicolumn{6}{c}{ TW~Hya} \\
\hline
CO 6--5        & $<3.22$   & $<1.19$  &2.46 &14.5       &CSO &Jun '00   \\
CO 4--3$^{d}$        & 5.0   & &            &11.0       &JCMT&     \\
CO 3--2        & 1.98     & 2.94     &0.63  &13.8       &JCMT&Nov '99\\
CO 3--2        & 1.00  & 0.77     &1.23     &25.7       &CSO &Jun '00\\
CO 2--1$^{d}$        & 1.02&         &      &20.0       &JCMT&       \\
$^{13}$CO 3--2 & 0.24     & 0.29     &0.78  &14.4       &JCMT&Feb '99\\
$^{13}$CO 2--1$^{d}$ & 0.14    &&           &20.0       &JCMT    &   \\
HCO$^{+}$ 4--3 & 0.49     & 0.72     &0.63  &13.4       &JCMT& Nov '99\\
H$^{13}$CO$^{+}$4--3& 0.07& 0.08     &0.76  &13.7       &JCMT& Dec '99\\
 HCN 4--3 &  0.49  &  0.60   & 0.76  & 13.5 & JCMT&  Dec '00\\
 H$^{13}$CN 4--3 &  $<$0.04$^{f}$  &  $<$ 0.05$^{c}$   & 0.76  & 13.5 & JCMT&  Dec '00\\
HCN 3--2$^{d}$ & 0.45      & &              &20.0       &JCMT&   \\
\hline
\end{tabular} \\
$^{a}$ Width of best single gaussian fit to total profile\\
$^{b}$ Calculated assuming the line is
double peaked consisting of two separate gaussians, each with a width
of 1.3 km s$^{-1}$ \\ 
$^{c}$ Listed value is 2$\times$rms \\ 
$^{d}$ Values from Kastner et al. (1997)\\ 
$^{e}$ Values from \cite{Qi} \\
$^{f}$ Calculated assuming a line width of 0.76 km s$^{-1}$ \\ 
\label{tab: nr1}
\end{table*}
\end{center}

\section{Simple one-dimensional analysis}
\label{sec:  radex}

Although the observed line intensities are a complex function of the
physical structure of the disk and the line/continuum optical depth,
useful insights can be obtained from a simple one dimensional analysis
of the line ratios. For constant temperature and density models such
as presented by Jansen et al.\ (1994) and Jansen (1995), the data
provide constraints on both parameters. To compare data obtained with
different beams the intensities were scaled to the same beam (see \S 4.3).

Consider first the observed ratios of $^{12}$CO and its
isotopomers. The $^{12}$CO 3-2/$^{13}$CO 3-2 ratios of 3.3 and 7.6 for
LkCa~15 and TW~Hya, respectively, indicate that the CO lines are
optically thick, assuming a normal isotope ratio of
[$^{12}$C]/[$^{13}$C]=60 in the solar neighborhood.  On the other
hand, the $^{13}$CO emission has an optical depth of only a
few, since the C$^{18}$O 3--2 emission is not detected. The ratio of
peak 3--2 temperatures of $>$3 (using the 2$\sigma$ limit for
C$^{18}$O) and the observed beam-corrected 1--0 ratio of 5.0 are only
marginally smaller than the isotope ratio [$^{13}$C]/[$^{18}$O] of
8.3 (Wilson \& Rood 1994).

The temperature can be determined from the $^{13}$CO 3--2/1--0 ratio
of 1.35$\pm$0.4, which gives temperatures of $\sim$20--40 K in
LkCa~15. Care should be taken with the interpretation of this result since
the emission of the two lines most likely comes from different regions
of the disk due to the difference in optical depth of the two lines
(see \S \ref{sec: tau1}).  The beam-corrected ratio for $^{13}$CO
3--2/2--1 of 0.9 for TW~Hya indicates that the bulk of the gas in this
source is colder than 25 K for densities $>10^{5}$ cm$^{-3}$.  The
$^{12}$CO 6--5 line probes higher temperatures since its upper level
lies at an energy of 116 K. The observed ratio $^{12}$CO
6--5/3--2=0.42$^{+0.21}_{-0.14}$ for LkCa 15 also
suggests the presence of gas with a
temperature
in the range 20--40 K while the upper limit 
of CO 6--5/3--2 $<1.0$ for TW~Hya gives $T<$150 K (cf.\
Figure 4 of \cite{jansen3}).  The 4--3 (JCMT) / 3--2 (CSO) ratio of
0.91$^{+0.45}_{-0.31}$ suggests temperatures $\sim$40~K whereas
the ratio of both JCMT lines indicates somewhat higher temperatures.  The
different optical depths of the 3--2 and 6--5 lines imply that they
probe different vertical layers of the disk.  Such vertical structure
may affect these conclusions, though not by large factors (see \S
\ref{sec: tau1}).

The density is best probed by molecules with a large dipole moment
such as HCO$^+$ and HCN. The measured HCO$^+$/H$^{13}$CO$^+$ 1--0
ratio of 6.2 for LkCa~15, and the 4--3 ratio of 6.7 for TW~Hya and
$>$1.9 for LkCa~15 indicate that the main isotopomeric lines are again
optically thick. No limits on HC$^{18}$O$^+$ exist, but H$^{13}$CO$^+$
may be close to optically thin (see \S \ref{sec: tau1}).  The limit on
the H$^{13}$CO$^+$ 4--3 line toward LkCa~15 together with the
1--0 line detected with OVRO gives a 4--3/1--0 ratio of
less than $<$ 2.4  and
constrains the density to $< 10^7$ cm$^{-3}$ in the HCO$^+$ emitting
region.  The optically thick HCO$^+$ 4--3/1--0 ratio of
0.75$^{+0.38}_{-0.25}$ suggests $n > 10^6$ cm$^{-3}$ at $T=$20--30 K.
The HCN 1$_{2}$--0$_{1}$ to H$^{13}$CN 1$_{2}$--0$_{1}$ ratio of 1.4
indicates that both lines are severely optically thick. The HCN 4--3
line has an even higher critical density than that of HCO$^+$ 4--3.
The observed HCN 4--3/1--0 ratio of 1.0$^{+0.5}_{-0.3}$ indicates
densities $n\approx 10^{7}-10^{8}$ cm$^{-3}$ in the LkCa 15 disk. 

For the TW~Hya disk,  the HCN/H$^{13}$CN 4--3 ratio has a lower
limit of 12.3, indicating that the HCN lines are optically thin or
nearly optically thin. The HCN 4--3/3--2 ratio of 0.6$^{+0.3}_{-0.2}$
constrains its density to lie in the 10$^6$ to 10$^8$ cm$^{-3}$ range,
and, as the lines are (nearly) optically thin, this should refer to the
regions in the disk where HCN is most abundant.

In summary, the simple analysis indicates that the main isotope lines
are optically thick, but that the lines of $^{13}$C isotopomers of CO
and HCO$^+$ have at most moderate optical depths. The bulk of the gas
is cold, but the presence of a warm layer is suggested from 
the CO 4--3 line for TW Hya. The inferred densities in the
region where the lines originate are high, at least $10^6$ cm$^{-3}$,
but not sufficient to  thermalize all transitions, especially those
from high dipole moment species.
\section{Description of models}
\label{sec: model}

\subsection{Adopted disk models}

\begin{figure}
\resizebox{\hsize}{!}{\includegraphics{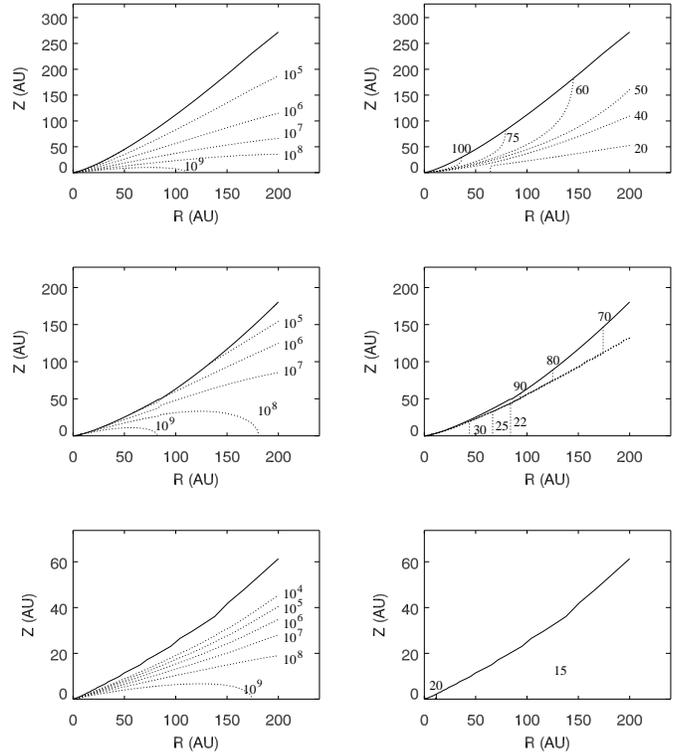}}
    \caption{\footnotesize Comparison of the density in cm$^{-3}$
(left three figures)
    and the temperature in K (right three figures) for the models
    of D'Alessio et al.\ (1999) (top), Chiang \& Goldreich (1997)
    (middle) and Bell (1999) (bottom). All models have a gas+dust mass of 0.024
    M$_{\odot}$. Only one quadrant of the 2-D flaring disk is
    shown.}
\label{fig: his1}
\end{figure}

In this work, the line emission from three recently published disk
models is calculated and compared with observations.  Although each
of these models has limitations, they are representative of the range
of temperatures and densities that may occur in disk models, even if
not specifically developed for the large radii probed in this work.
The three disk treatments analyzed in detail are:
\begin{enumerate}  

\item{The model by D'Alessio et al.\ (1999) (see also D'Alessio,
Calvet \& Hartmann 1997 and D'Alessio et al. 1998), in which
the disks are geometrically thin and in vertical hydrostatic
equilibrium. The gas and dust are heated by viscous dissipation,
radioactive decay, cosmic rays and stellar irradiation. The gas and
dust are coupled and can thus be described by a single
temperature. The assumption of a geometrically thin disk implies that
the energy balance can be calculated at each radius separately,
decoupled from other radii. The temperature and density distribution
(Fig. \ref{fig: his1}, top two Figs.) have been calculated using this
procedure.}

\item{ The model by Chiang \& Goldreich (1997, 1999) represents a
passive disk in radiative and hydrostatic balance.  The disk structure
is bi-layered with a super-heated dust surface layer that is in
radiative balance with the light from the central star. This
super-heated layer radiates half of its energy toward the midplane,
thereby heating the inner regions of the disk. Figure \ref{fig: his1}
(center two figures) presents the temperature and density distribution
from this model.  
} 

\item{The model by Bell et al.\ (1997) and Bell (1999) takes the
effect of both stellar irradiation and the reprocessing of radiation
into account. This redistribution of energy gives rise to a strong
flaring of the disk near the star, but less strongly further away. The
assumption of vertical hydrostatic equilibrium is used; however, the
flaring of the disk at small radii (the first few AU) may, for high
values of the mass accretion rate $\dot{M}$, shield the surface of the
disk at larger radii from the radiation. The resulting temperature and
density distribution is shown in Figure \ref{fig: his1} (bottom two
figures).  The shielding of the outer regions results in a cold disk
which is approximately isothermal in the vertical direction. }
\end{enumerate}

The published models described above were calculated for different
disk masses. Since the results depend sensitively on mass and cannot
simply be scaled, this greatly complicates their comparison.  To
eliminate this uncertainty, the authors have kindly supplied several
of their model runs for various masses. The model presented here for
LkCa~15 has a mass of 0.024 M$_{\odot}$, close to its observed value
(\cite{1995ApJ...439..288O}). The mass of the TW~Hya disk estimated
from submillimeter and centimeter continuum observations is $\sim$0.03
M$_{\odot}$, assuming a gas/dust ratio of 100/1 (Holland et al.\ 2000,
Wilner et al.\ 2000). The disks all have an outer radius of 200 AU,
and the precise mass is fixed by dividing the density of an
appropriate model by a small factor of less than 3. Figure \ref{fig:
histo} compares the fraction of mass at a given temperature or density
in the three models, whereas the distributions within the disk are
shown in Figure \ref{fig: his1}. While the density distributions are
similar, the temperature distributions between the three models are
clearly different. One of the aims of this paper is to investigate
whether the molecular line data are consistent with these different
types of models.

 The adopted models are not tailor-made for the two sources
studied here.  For example, they do not fit in detail the observed
SEDs: the Bell (1999) model is too cold on the outside to reproduce
the mid-infrared emission.  Chiang et al.\ (2001) have presented
models for LkCa 15 and TW Hya which fit the observed SEDs. However,
both models have significant setting of the dust.  The gas may still
flare out to higher vertical distances, but the SED does not provide
observational constraints. For this reason, we adopted the original
Chiang \& Goldreich (1997) model which has no settling of dust so that
the temperature is defined over the entire disk. Other parameters
entering the models are the disk accretion rate and the luminosity and
effective temperature of the star. For the accretion rate, which
enters the D'Alessio et al.\ models, a value of 10$^{-8}$ M$_{\odot}$
yr$^{-1}$ and $\alpha=0.01$ was chosen, which is higher than the
observed values of 10$^{-9}$ and $5\times 10^{-10}$ M$_{\odot}$
yr$^{-1}$ for LkCa 15 and TW Hya, respectively (Hartman et al.\ 1998,
\cite{2000ApJ...535L..47M}). However, the observed molecular lines probe
the outer region of the disk whereas the accretional heating due to
the high values of $\alpha$ and the accretion rate will only affect
the inner few AU. The models refer to a 0.5 M$_{\odot}$
star with T$_{\mbox{eff}}$=4000 K. For comparison, LkCa~15 has a mass
of roughly 1 M$_{\odot}$ and an effective temperature of 4400 K (Siess
et al.\ 1999), while TW Hya has a mass of 0.7 M$_{\odot}$ and
T$_{\mbox{eff}}$= 4000 K (\cite{2000ApJ...535L..47M}).  

\begin{figure}
\resizebox{\hsize}{!}{\includegraphics{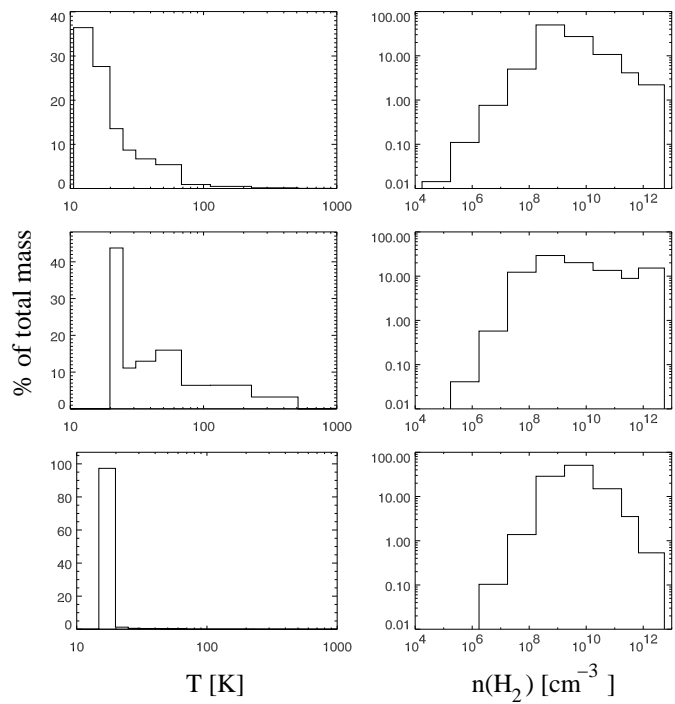}}
    \caption{\footnotesize Comparison of the fraction of mass in a
    given temperature and density interval for the models of D'Alessio
    et al.\ (1999) (top figures), Chiang \& Goldreich (1997) (middle
    figures) and Bell (1999) (lower figures). The distributions in the
    disks are plotted in Figure \ref{fig: his1}. }
\label{fig: histo}
\end{figure}

\subsection{Radiative transfer methods}

 The radiative transfer in the molecular lines from disks is
calculated in two steps. First, the abundances of the molecules in the
disk are estimated using a ray-tracing method in the vertical
direction through the disk and adopting statistical equilibrium
calculations. The ratio of the different modeled lines constrains the
range of depletions.  This calculation does not take into account the
inclination of the source and assumes that the ratio of different
lines is less sensitive to inclination than the integrated intensity
of a single line.

Once the depletions are constrained, a full 2D radiative transfer
calculation is performed using the accelerated Monte Carlo (AMC) code
of Hogerheijde \& van der Tak (2000), whose results are compared to
the observations. The motivation for this elaborate scheme is given in
\S \ref{sec: popul} and is driven by the huge computational time
involved in the latter calculation.  The AMC code has been compared
with other radiative line transfer codes in a workshop in Leiden in
1999, where the populations of the levels and convergence have been
tested for a set of one-dimensional problems. The comparison of the
various codes is described in van Zadelhoff et al.\ (in preparation).
\footnote{\footnotesize see also
http://www.strw.leidenuniv.nl/$\sim$radtrans}.

\subsection{Level populations and depletions}
\label{sec: popul}
\begin{figure}
\resizebox{\hsize}{!}{\includegraphics{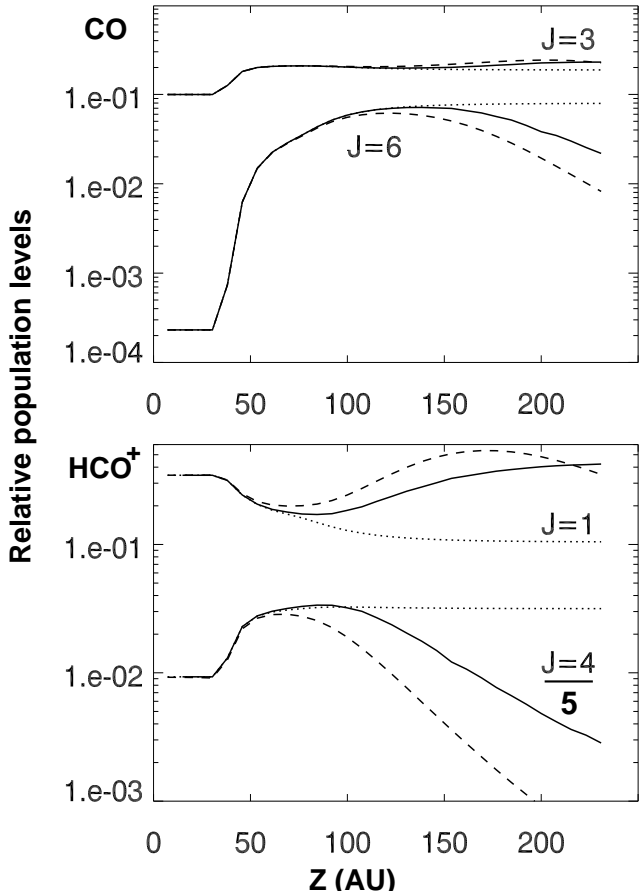}}
\caption{\footnotesize The relative populations of levels of the CO
molecule ($J$=3 and $J$=6) and the HCO$^{+}$ molecule ($J$=1 and
$J$=4) for three different calculations (LTE (dotted), SE with no
stimulated emission or absorption (dashed) and SE with full 2D
radiative transfer (solid)). The levels are calculated for the
D'Alessio et al. (1999) model and plotted as a function of height $Z$
at a radius of 175 AU. The assumed abundances are 10$^{-4}$ for CO and
5$\cdot10^{-9}$ for HCO$^{+}$. In the lower plot the HCO$^{+}$ $J$=4
level population is reduced by a factor of 5 for clarity. It is
clearly seen that the populations in the upper layer of the disk are
not in LTE. }
\label{fig: levelpop}
\end{figure}
The level populations $x_u$ and $x_l$ can be calculated in various
ways. The simplest approximation is that of Local Thermodynamic
Equilibrium (LTE), which is valid for all levels that are
collisionally excited in a gas with densities higher than the critical
density for that level. The latter is given by

\be
n_{cr}=\frac{A_{ul}}{\sum_{i} K_{ui}} \, {\rm cm^{-3}}, 
\end{equation} 

where $\sum_{i} K_{ui}$ denotes the sum of all collisional rate
coefficients from level u to all other levels i. LTE is usually
adopted in the analysis of line emission from disks (e.g., G{\'o}mez
\& D'Alessio 2000, Dutrey \etal \ 1996, Kastner et al. 1997). Although
the bulk of the gas is in LTE due to the high densities, the radiation
from the central part of the disk can be highly optically thick
depending on the molecular abundances and assumed depletions. In that
case the outer layers dominate the emission where the densities can
fall below the critical densities of the higher frequency lines for
high dipole moment molecules.  For these regions statistical
equilibrium (SE) calculations, also referred to as non-LTE or NLTE,
need to be performed.

 The populations of the levels are calculated by solving the equation:
\begin{equation}
\begin{split}
\sum_{j>l}[n_{j}A_{jl}+(n_{j}B_{lj}-n_{l}B_{lj})\overline{J}_{jl}] \\
-\sum_{j<l}[n_{l}A_{lj}+(n_{l}B_{lj}-n_{j}B_{jl})\overline{J}_{lj}] \\
+\sum_{j}[n_{j}C_{jl}-n_{l}C_{lj}]=0, \\
\end{split}
\label{eq: SE} 
\end{equation}
where $\overline{J}_{jl}$ is the integrated mean intensity, $A_{ij}$
and $B_{ij}$ the Einstein coefficients and $C_{ij}$ the collisional
rates.  The Einstein $A$ coefficients and collisional rate
coefficients for CO, HCO$^+$ and HCN are the same as those in Jansen
et al.\ (1994, Table 4).  The calculation of the level populations is
an iterative process since the integrated mean intensity is directly
related to the levels $n_{l}$, which in turn affects the mean
intensity.

Test calculations have been performed for three cases. The first is
LTE, where the populations are given by the Boltzmann equation. In
this assumption the populations are dominated by collisions and
therefore depend only on the local temperature. The second is
Statistical Equilibrium without stimulated radiative effects
(SE[$\overline{J}_{\nu}$=0]), in which the populations of the levels
are no longer assumed to be dominated by collisions and are calculated
explicitly. In this case, Equation~(\ref{eq: SE}) is solved under the
assumption that $\overline{J}_{\nu}$=0 for all radiative
transitions. The third method is the full Statistical Equilibrium (SE)
solution using a Monte Carlo code (\cite{hogerheijde}) to calculate
the mean intensity $\overline{J}_{\nu}$ for each radiative transition
iteratively, taking the line emission and absorption throughout the 2D
disk into account.

In Figure \ref{fig: levelpop}, the relative populations for the levels
$J$=3 and 6 of CO and $J$=1 and 4 of HCO$^{+}$ are plotted for the
LTE, SE($\overline{J}_{\nu}$=0) and SE for the D'Alessio et al.\
(1999) model. This model is chosen because it has a smooth temperature
gradient but does show a temperature inversion in the $z$-direction.
The adopted CO and HCO$^{+}$ abundances are 10$^{-4}$ and 5$\cdot
10^{-9}$ respectively, and the turbulent line width is assumed to be
0.2 km s$^{-1}$.  Figure \ref{fig: levelpop} shows that the
differences between the three methods are small in the midplane but
that they become significant in the lower density upper layers.  For
the $J$=1 level of HCO$^{+}$, the influence of the Cosmic Microwave
Background 2.7~K radiation is apparent since its relative population
continues to rise toward the outside in the SE calculation compared to
the SE($\overline{J}_{\nu}$=0) calculation.
 
Even though only the full 2D SE calculation describes the populations
accurately, its calculation is an enormous computational task due to
the large column densities and steep gradients in density and
temperature coupled with a narrow velocity profile.  In these cases,
the convergence criteria of a numerical code become very important.
The SE($\overline{J}_{\nu}$=0) calculation provides better agreement
with the SE calculation compared to LTE, especially at larger
distances from the star where most of the observed radiation
originates. Therefore the SE($\overline{J}_{\nu}$=0) method is adopted
in the ray-tracing calculations.

\begin{figure}
\resizebox{\hsize}{!}{\includegraphics{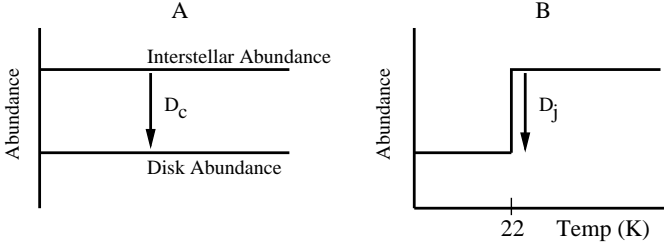}}
\caption{\footnotesize The two ways in which the depletion of molecules
compared to the interstellar value is taken into account in the
computations. On the left a constant depletion $D_{C}$ is shown,
whereas on the right a step-function representing the freeze-out of
molecules $D_{J}$ below a critical temperature is illustrated}
\label{fig: deplet}
\end{figure}

The abundances and depletion of various molecules are taken into
account in two different ways:
\begin{enumerate} 
\item{A constant model, in which the fractional abundances are the
same throughout the disk. In the standard model, abundances close to
those found in dark clouds are chosen.  These values can be
subsequently lowered by a constant factor $D_C$ with respect to the
standard values. Note that the chemical interpretation of $D_C$ may
vary: it can be due to freeze-out, photodissociation, different
chemistry at high or low densities, or any combination of these. A
sketch of the depletion models compared to interstellar abundances is
given in Figure \ref{fig: deplet}.}
\item{A jump model, in which the abundances are constant except for
regions where the temperature is lower than a critical
temperature. Below this temperature the abundance is assumed to drop
by a certain factor due to freeze-out of the molecule onto cold dust
grains. In this paper, a temperature of 22 K (based on CO freezing
onto a CO surface; \cite{1993ApJ...417..815S}) is assumed (see Figure
\ref{fig: deplet}). HCO$^{+}$ is assumed to follow the CO abundance
profile and will thus deplete at the same temperature. For HCN a
critical temperature of 80 K is assumed. A more detailed and realistic
description of the depletion and abundance variations has been given
by Aikawa \& Herbst (1999) and Willacy et al. (1998), but these
results are too specific for the exploratory purposes of this paper.}
\end{enumerate}

\subsection{Approximate line intensities}
\subsubsection{Vertical radiative transfer}
\label{sec: basic}
 To constrain the depletions and thereby bracket the computational
domain of the problem, the intensity of each line is calculated by
solving the radiative transfer equation \be
\frac{dI_{\nu}}{ds}=\alpha_{\nu}(\frac{j_{\nu}}{\alpha_{\nu}}-I_{\nu})~~~~~,
\end{equation} in the vertical direction, where $I_{\nu}$ is the intensity, $s$
the path length along a ray normal to the disk, $j_{\nu}$ the emission
function, and $\alpha_{\nu}$ the absorption function, which is the
inverse of the mean free path. The ratio of the emission and
absorption functions is known as the source function.
The transfer equation is solved in an iterative ray-tracing procedure
using small $\Delta s$ steps from $s_{0} = -\infty$ to $s_{N}=
+\infty$ (the observer).

The equation to be solved thus becomes
\be I_{\nu ,{\rm line}}(s_{i})=I_{\nu}(s_{i-1}) e^{-\tau_{\nu}} +
j_{\nu_{0}} (s_{i}) \phi(\nu) \frac{(1-e^{-\tau_{\nu}})}{\tau_{\nu}}
\Delta s_{i}~~~~~,
\label{eq: ray}
\end{equation} where $\phi(\nu)$ is the line emission profile function in
frequency space and $I_{\nu}(s_{0})=0$. 

The optical depth $\tau_{\nu}$ is the sum of the attenuation by dust
and gas along the step-length and is equal to \be \tau_{\nu} =
(\alpha_{c}+\alpha_{l}) \, \Delta s~~~~~, \, \end{equation} where \ba \alpha_{c}
& = & \kappa_{\nu}\frac{m_{\rm gas}}{R_{gd}}\,n_{\rm gas}~~~~~\, , \\
\alpha_{l}& = & \frac{A_{ul} c^{3}}{8 \pi \nu^{3}}
\left(x_{l}\frac{g_{u}}{g_{l}}-x_{u}\right) n_{\rm gas} X_{m}
\phi(\nu)~~~~~,
\ea
and the emission is given by
\begin{equation}
j_{\nu} =\frac{h \nu}{4 \pi}A_{ul}x_{u}n_{\rm gas} X_{m}\phi(\nu).
\end{equation}
In these equations, $m_{gas}$ is the mean
molecular weight of a gas particle, $R_{gd}$ the gas to dust ratio,
$A_{ul}$ the Einstein coefficient, $\nu$ the frequency of the
transition, $c$ the speed of light, $X_{m}$ the abundance of the
molecule relative to the gas density $n_{gas}$, which is taken to be
equal to the density of H$_{2}$, $g_i$ the statistical weight of level
$i$ and $x_{i}$ the population of level $i$.  A mean molecular weight
of 2.2 (proton mass per particle) and $R_{gd}=100$ are adopted.  The
level populations are solved using SE ($\overline{J_{\nu}}$=0) (\S \ref{sec:
popul}) for the reasons stated above.

The continuum mass absorption coefficient $\kappa_{\nu}$ is taken
from  \cite{ 1994A&A...291..943O} and extended to wavelengths
longer than 1.3 mm as

\be \kappa_{c}=\kappa_{c}(1.3\mbox{mm}) \left(\frac{\nu}{2.31 \times
10^{11}\ {\rm Hz}} \right)^\beta~~~~~,
\end{equation} where $\kappa_{c}(1.3 \, \mbox{mm})$
depends on the specific coagulation model chosen 
 and $\beta$ is taken to be 1. For this problem the $\kappa_c$
values were taken from the model with coagulation at a density of
10$^{8}$ cm$^{-3}$ covered by a thin ice-layer ($\kappa_{c}$(1.3
mm)=1.112 cm$^{2}$ g$^{-1}$). In practice, however, the
dust absorption is negligible compared to the line absorption for the
molecular transitions in the wavelength range of interest.

\subsubsection{Calculation of intensity}
\label{sec: basic2}
For comparison with observations and model results, all values are
referred to the size of the disk model. The observations are thus
scaled as follows:

\ba T_{\rm disk,\mbox{obs}}&=&\frac{T_{A}^{\ast}}{\eta_{\mbox{tel}}}
\frac{\Omega_{\nu,\mbox{tel}}}{\Omega_{\nu,\mbox{disk}}}\, \\
\Omega_{\mbox{disk}}&=&\frac{2 \pi
\int_{R_{\rm min}}^{R_{\rm max}} R dR}{{\rm AU}^{2}} 
\frac{1}{(d ({\rm pc}))^{2}} \,~~~~~,
\ea
\noindent
where the ratio $T_{\rm mb}={T_{A}^{\ast}}/{\eta_{\mbox{tel}}}$ is given in
Table \ref{tab: nr1}, $\Omega_{\nu,\mbox{tel}}$ is the beam size at
the frequency of the line in arcsec$^{2}$, $d$ the distance in parsec, AU the
astronomical unit, and $\Omega_{\mbox{disk}}$ the size of the disk in
arcsec$^{2}$.

The models are calculated using the ray-tracing Equation (\ref{eq:
ray}) which are weighted according to its emitting surface area  
\be
\overline{I}=\frac{\sum_{i=0}^{Nx} I_{i} \pi
(R_{i}^{2}-R_{i-1}^{2})} {2 \pi \int_{R_{\rm min}}^{R_{\rm max}}
R dR}. 
\label{eq: scale}
\end{equation} Here, $\overline{I}$ is the mean intensity for the entire disk at
the surface, with Nx the number of cells in the $R-$direction.  The
disk is assumed to be seen face-on.  This is a very good approximation
for TW~Hya, whereas it should still be reasonable for LkCa~15 even
though the disk is seen at an inclination of 60\degr. The intensities
derived by this method will be used only in the analysis of line
ratios, which are less sensitive to inclination effects than absolute
values.

 The radius of the disk is taken to be 200 AU in all models, or 400 AU
diameter.  The size of the LkCa~15 disk ($d$=140 pc) suggested by the
OVRO $^{13}$CO maps of Qi (2000) is slightly larger (420 AU$\times$530
AU). For TW~Hya, no millimeter interferometer observations are
available, but mid-infrared and VLA 7 millimeter images suggest a disk
size of $\sim$100 AU (\cite{wilner}). Scattered light images observed
with the {\it Hubble Space Telescope} suggest an outer radius of at
least 200 AU, however (Weinberger et al. 1999, Krist et al. 2000). We
therefore adopt a similar disk size in AU as for LkCa~15, but with
$d$=57 pc.

\subsection{Line intensities using full radiative transfer}
 For the calculation of the populations in SE using full radiative
transfer, the Monte Carlo code developed by Hogerheijde \& van der Tak
(2000) is used. In this code, Equation (\ref{eq: SE}) is solved in an
iterative fashion, where all photons start at the outer boundary with
an intensity given by the 2.728~K Cosmic Background radiation.  In
this calculation, the inferred abundances from the
SE($\overline{J_\nu}$=0) method are adopted.  The calculated
populations at each position in the disk are used to compute the
complete line profiles of selected molecules using a program which
calculates the sky brightness distribution.  The profiles are
calculated by constructing a plane through the origin of the disk
perpendicular to the line of sight, with a spatial resolution small
enough to sample the physical and velocity distributions.  Both the
spatial resolution and the velocity resolution can be specified. A
ray-tracing calculation is performed through this plane from $-\infty$
to $+\infty$, keeping track of the intensity in the different velocity
bins.  

\section{Results} 
\label{sec: results}

\begin{figure*}
\resizebox{\hsize}{!}{\includegraphics{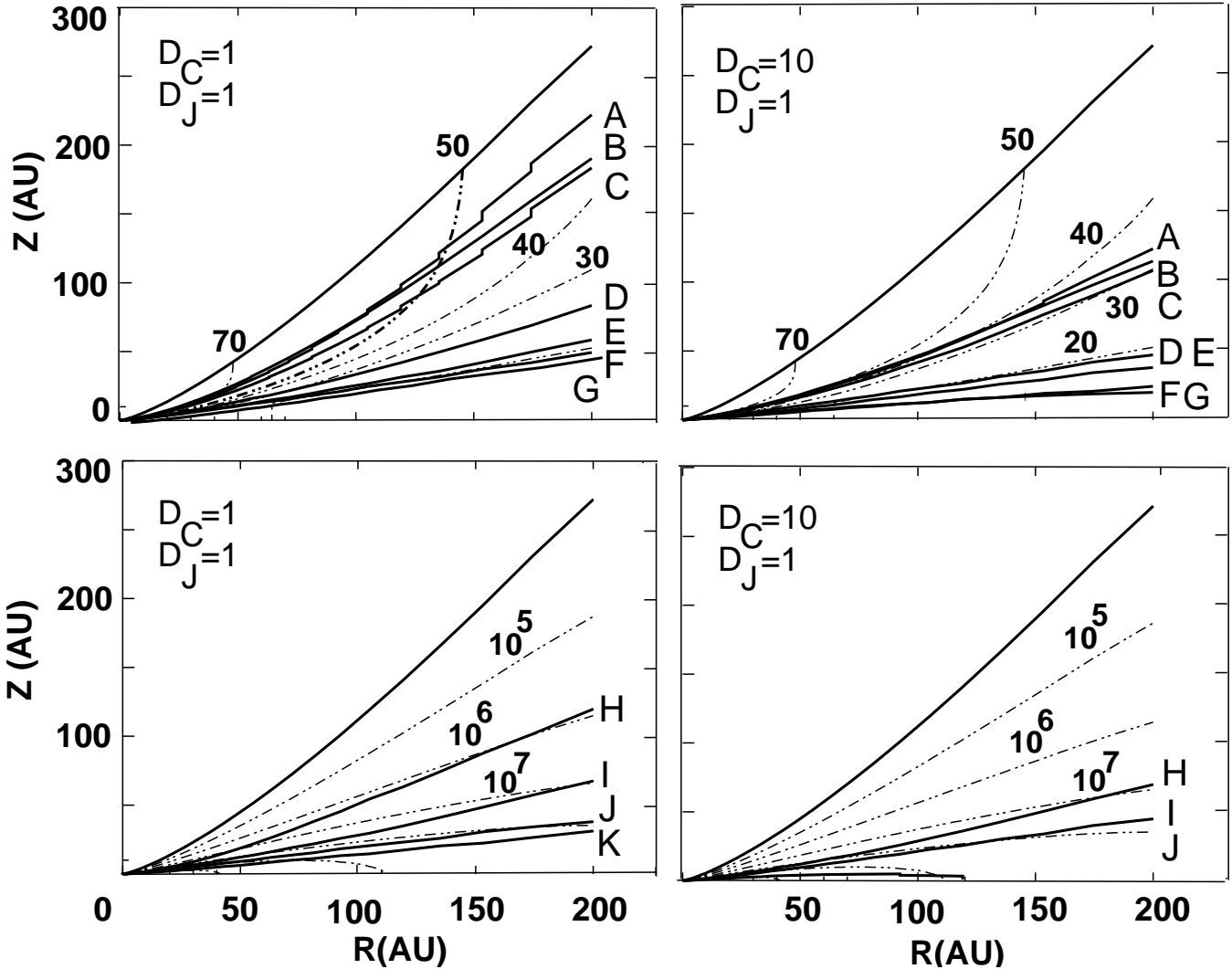}}
\caption{\footnotesize The $\tau=1$ surfaces for the observed CO and
HCO$^{+}$ isotopomer lines integrated from the top, overplotted on the
temperature (top) and density distribution (bottom) in the model by
D'Alessio et al.\ (1999). The dotted contours are iso-temperature 
(in ~K) or
iso-density (in cm$^{-3}$) contour lines.  The results are shown for both the
standard abundances (left) and depleted by a constant factor $D_C=10$
(right). The labels indicate: A: $^{12}$CO 6--5; B: $^{12}$CO 3--2;
C: $^{12}$CO 2--1; D: $^{13}$CO 3--2; E: $^{13}$CO 1--0; F: C$^{18}$O
3--2; G: C$^{18}$O 2--1; H: HCO$^+$ 4--3; I: HCO$^+$ 1--0;
J: H$^{13}$CO$^+$ 4--3; K: H$^{13}$CO$^+$ 1--0.
}
\label{fig: temp}
\end{figure*}

The models are calculated initially using standard dark cloud abundances
of CO of 1$\cdot$10$^{-4}$, HCO$^{+}$ of 5$\cdot$10$^{-9}$, and HCN of
5$\cdot 10^{-9}$ relative to H$_{2}$. For the isotope ratios the
following values are used throughout:
${^{12}\mbox{C}}/{^{13}\mbox{C}}$=60 and
${^{16}\mbox{O}}/{^{18}\mbox{O}}$=500.  These models are referred
to as $D_C$=$D_J$=1. Subsequently, the values of 
$D_C$ and $D_J$ are varied (see \S 4.3).
The line
intensities have been calculated assuming a micro-turbulence of 0.2 km
s$^{-1}$. For TW~Hya, this results in calculated line widths
of $\sim 0.8$ km s$^{-1}$, in good agreement with observations.

\subsection{$\tau=1$ surfaces}
\label{sec: tau1}

Significant insight into the observational results can be obtained by
investigating the regions of the disk where the different lines become
optically thick.  At each radius the effective emission region for
each line is calculated using the SE($\overline{J_\nu}$=0) method by
integrating from the top layer down until $\tau$=1 in line + continuum
is reached. Although the $\tau$=1 level is chosen arbitrarily and
radiation from deeper in the disk may still escape, it provides a
useful measure of the volume of the emitting region for each molecular
transition.  This calculation is performed only for a face-on disk for
simplicity and is thus only applicable for the TW~Hya case. It does,
however, give an indication of the parts of the disk from which the
molecular emission arises in more general cases.  For the model by
D'Alessio et al.\ (1999), a contour-plot of the $\tau=1$ surfaces of
the observed CO and HCO$^+$ lines is given in Figure \ref{fig: temp},
where the former are overplotted on the temperature distribution and
the latter on the density distribution.  The line emission is
dominated by densities and temperatures above the $\tau=1$ contour,
which can then be compared to the values derived from the constant
temperature and density models given in \S \ref{sec: radex}.

It is seen that, for standard abundances, the $^{12}$CO lines become
optically thick in the upper, warm layer of the disk where $T>40$~K.
On the other hand, the $^{13}$CO lines probe into the colder regions
around 20--30~K. Thus, the $^{12}$CO excitation temperature should be
higher than that of $^{13}$CO, which must be taken into account in the
analysis of isotopomeric line ratios.  Similarly, the higher frequency
3--2 and 6--5 lines generally have higher optical depths than the 1--0
lines, and thus probe better the warmer upper layer. Even C$^{18}$O is
not fully optically thin, but has $\tau\approx 1-2$.
The low temperature of 20--30~K probed by $^{13}$CO is consistent
with the simple analysis of the data in \S 3.

For the standard HCO$^+$ abundance, the 1--0 to 4--3 lines are optically
thick in the outer layers, whereas the H$^{13}$CO$^+$ lines are close
to optically thin throughout the disk. Thus, the HCO$^+$ lines probe
densities of order $10^6 - 10^7$ cm$^{-3}$, below the critical density
of the 4--3 transition. For H$^{13}$CO$^+$, the populations will be
closer to LTE because its emission arises primarily from regions with
densities of $10^7-10^8$ cm$^{-3}$.  If the HCO$^+$ abundance is
depleted by a constant factor $D_C$=10, the HCO$^+$ lines
become optically thin in the outer regions of the disk and now trace regions
with densities of $10^7-10^8$ cm$^{-3}$.  The HCN 1--0 to 4--3 lines
show a similar behavior to HCO$^+$.  The densities of $10^6 - 10^8$
cm$^{-3}$ derived from the observed HCO$^+$ and HCN lines in \S 3 are
consistent with this analysis for modest depletions of both species.

\subsection{Integrated line ratios: range of depletions}
\label{sec: lineratio}

\begin{table*}[Ht]
\begin{center}
\caption{Ranges of inferred depletions 
for the different molecules for the three disk models}
\begin{tabular}{llcccccc}
\hline
&   & \multicolumn{2}{c}{D'Alessio et al.\ (1999)}
&\multicolumn{2}{c}{Chiang \& Goldreich (1997)} 
&\multicolumn{2}{c}{Bell (1999)} \\   
 &    & LkCa~15         &TW~Hya &LkCa~15    & TW~Hya & LkCa~15 & TW~Hya \\
\hline
CO       &$D_C$& [3,15]&$>$30&[3,30]&$>100$ &[1,500] & [10,1000] \\
 &$D_J$& [3,30]&$>$1&[1,15]& $>$1 &$>$1 &$>$1 \\
HCO$^{+}$&$D_C$& [3,80]&$>80$&[10,100]&$>100$&[1,80] &  [2,1000]  \\
 &$D_J$& $>$1 &$>$1&$>$10&$>$1&[1,100]&$>$1\\
HCN      &$D_C$& [1,400]&[4,600]&[10,200]&[10,800]  &[1,500] &  [4,500]   \\
 &$D_J$& $>$1 &$>$1&$>$1&$>$1&$>$1&$>$1       \\
\hline
\end{tabular} \\
The numbers in square brackets indicate the range of inferred depletions \\
\label{tab: tab2}
\end{center}
\end{table*}

In this section, the relative line intensities obtained in the
SE($\overline{J_\nu}$=0) method are used
to constrain the abundances of the molecules and the level of
depletion.  Since lines of different isotopomers arise from different
regions, their line ratios will depend on the local depletion values.
By calculating models for a range of depletions, the abundances 
can be derived by varying both the overall depletion $D_C$
and the jump depletion $D_J$ as described in \S \ref{sec: popul}.
In the comparison of the line ratios, the difference in beam dilution
for the two lines must be taken into account (\S \ref{sec: basic2}).
 
\begin{figure}[h] 
\resizebox{\hsize}{!}{\includegraphics{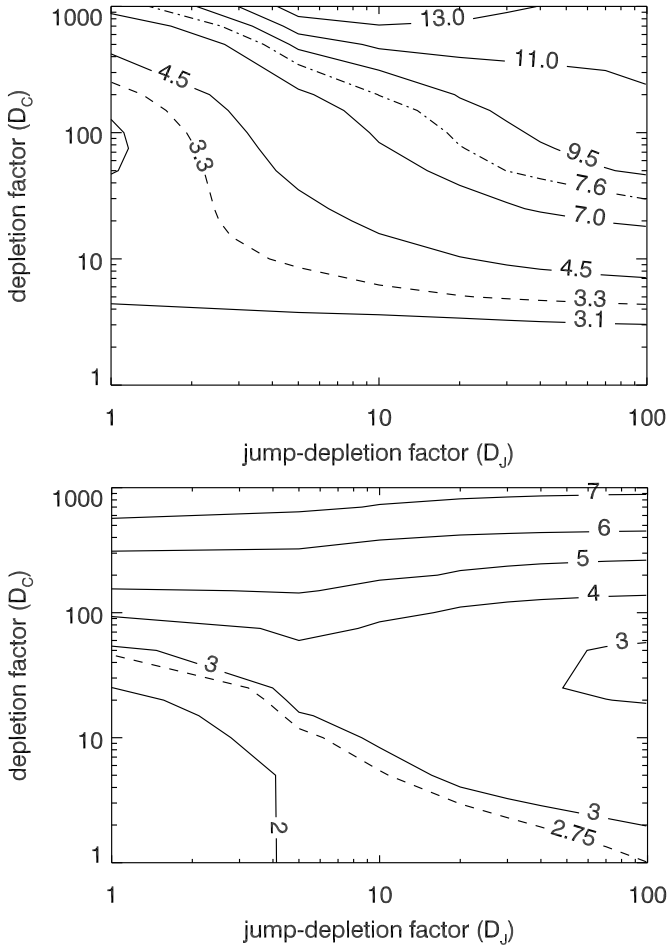}} 
\caption{The CO/$^{13}$CO 3--2 (top) and $^{13}$CO/C$^{18}$O 3--2
(bottom) line intensity ratios as functions of the jump depletion
$D_J$ and the overall depletion $D_C$ within the D'Alessio et al.\
(1999) model. The observed ratios for LkCa~15 CO 3--2/$^{13}$CO 3--2
(dashed line) and TW~Hya (dash-dotted line) are shown in the figures.}
\label{fig: ratio1} 
\end{figure}

Isotope ratios are more sensitive to both $D_C$ and $D_J$ as compared
to ratios of different species. For instance the $^{12}$CO emission
remains optically thick up to large values of $D_C$ and therefore does
not probe the region below 22~K, whereas $^{13}$CO becomes sensitive
to $D_J$ for modest values of $D_C$.  In Figure \ref{fig: ratio1} the
CO/$^{13}$CO 3-2 and $^{13}$CO/C$^{18}$O 3-2 line intensity ratios are
plotted as functions of $D_J$ and $D_C$.  The observed values for
LkCa~15 are plotted with dashed lines, and indicate a range $D_C \in
[3,40]$ when both plots are combined. The ratio for TW~Hya (dot-dashed
line) indicates a larger depletion with $D_C > 100 $ and $D_J \gtrsim
10$. Combining similar plots for all species and lines, the resulting
values of $D_C$ and $D_J$ are shown in Table \ref{tab: tab2} for the
three models of interest for both sources. The inferred ranges for the
disk models are large and it is difficult to give accurate values for
molecules for which few lines have been measured. The abundances of
all molecules in the TW~Hya disk seem to be depleted by a large
factor. In general $D_J \approx 10$ is taken as a best fit for both
sources. In cases where no constraints are available, $D_{J}=10$ has
been assumed.

\begin{table*}[Ht]
\caption{Integrated intensities for the higher rotational
lines for the three
disk models}
\begin{tabular}{llllllllllll}
\hline
\\[-0.30cm]
Model & D$_C$ & D$_J$ &
CO 6--5 & CO 3--2 & $^{13}$CO 3--2 & C$^{18}$O 3--2 & HCN 4--3 & H$^{13}$CN 4--3 & HCO$^{+}$ 4--3 & H$^{13}$CO$^{+}$ 4--3 \\
\hline
A$^a$ &1&1     &0.78   &1.15  &0.39   & 0.18 & 0.39 &0.12 &0.54 &0.15 \\
B$^b$ &1&1     &0.076 &0.21  &0.16   & 0.11 & 0.15 &0.092 &0.17 &0.10  \\
C$^c$ &1&1     &0.59   &0.80  &0.42   & 0.28 & 0.42 &0.20 &0.48& 0.24\\
A &5&10    &0.61   &0.93  &0.24   & 0.074 & 0.25 &0.030 &0.37 &0.043\\
B &5&10    &0.046  &0.16  &0.089   & 0.036 & 0.096 &0.011 &0.11 &0.018\\
C &5&10    &0.36   &0.61  &0.23   & 0.086 & 0.26 &0.049 &0.33 &0.062\\
A &10&10   &0.44  &0.68 &0.16   & 0.044 & 0.17 &0.018 &0.26 &0.026\\
B &10&10   &0.041 &0.15 &0.074   & 0.022 & 0.081 &0.006 &0.095 &0.009\\
C &10&10   &0.25  &0.46 &0.16   & 0.057 & 0.19 &0.031 &0.24 &0.041\\
\hline
\\[-0.3cm]
\multicolumn{3}{c}{LkCa~15$^{d}$} & 0.53& 1.39 & 0.39 & $<$0.14& 0.25 
& $\ldots$ & 0.26 & $<$0.13 \\
\hline
Model & D$_C$ & D$_J$ &
CO 6--5 & CO 3--2 & $^{13}$CO 3--2  & HCN 4--3& HCN 3--2 & H$^{13}$CN 4--3 & HCO$^{+}$ 4--3 & H$^{13}$CO$^{+}$ 4--3 \\
\hline
A &100&10& 0.88 &   1.84 &0.29  &0.36& 0.22 &0.024 &0.52 &0.038   \\
B &100&10& 0.12 &   1.07 &0.14  &0.18& 0.13 &0.004 &0.26 &7.0E-3  \\
C &100&10& 0.60 &   1.46 &0.36  &0.46& 0.24 &0.044 &0.58 &0.065   \\
A &200&10& 0.68 &   1.81 &0.23  &0.29& 0.17 &0.013 &0.42 &0.021   \\
B &200&10& 0.069&   0.95 &0.083 &0.11& 0.08 &0.002 &0.18 &3.4E-3  \\
C &200&10& 0.57 &   1.73 &0.32  &0.43& 0.21 &0.025 &0.55 &0.040 \\
\hline
B & 10&10& 0.38 &   1.41 &0.15  &0.54& 0.48 &0.045 &0.62 &0.06 \\
\hline
\multicolumn{3}{c}{TW Hya$^{d}$} & $<$3.2 & 1.98 & 0.24  
  & 0.49 & 0.45 &$<$0.04 & 0.49 & 0.07  \\
 \hline
\end{tabular} \\
$^a$ D'Alessio et al.\ (1999) model \\
$^b$ Bell (1999) model \\
$^c$ Chiang \& Goldreich (1997) model \\
$^{d}$ The observed values have an estimated uncertainty of 20 \%; all
values refer to the original beam size of the observations
(see Table \ref{tab: nr1})
\label{tab: tab3}
\end{table*}

\begin{table*}[Ht]
\caption{Integrated intensities for the lower rotational lines for the three
disk models}
\begin{tabular}{llllllllllll}
\hline
\\[-0.30cm]
Model & D$_C$ & D$_J$ &
$^{13}$CO 1--0 & C$^{18}$O 1--0 & HCN 1--0 & H$^{13}$CN 1--0 & HCO$^{+}$ 1--0 & H$^{13}$CO$^{+}$ 1--0 \\
\hline
A$^a$ &1& 1 &6.18&1.90&4.47&0.85&4.84&0.20 \\
B$^b$ &1& 1 &4.04&1.57&2.58&0.76&2.68&0.17 \\
C$^c$ &1& 1 &8.80&2.88&5.72&0.96&5.94&0.25 \\
A &5&10 &3.60&0.53&2.70&0.11&3.11&3.3E-2 \\
B &5&10 &2.04&0.32&1.39&5.5E-2&1.56&1.8E-2 \\
C &5&10 &3.56&0.62&2.52&0.14&3.06&4.0E-2 \\
A &10&10&2.31&0.31&1.75&6.0E-2&2.11&1.8E-2 \\
B &10&10&1.55&0.18&1.11&2.8E-2&1.31&9.3E-3 \\
C &10&10&2.40&0.38&1.70&7.6E-2&2.10&2.3E-2\\
\hline
\\[-0.3cm]
\multicolumn{3}{c}{LkCa~15$^{a}$} &7.43&0.70 &3.04&1.20&4.19&7.E-2 \\
 \hline
\end{tabular} \\
$^a$ D'Alessio et al.\ (1999) model \\
$^b$ Bell (1999) model \\
$^c$ Chiang \& Goldreich (1997) model \\
$^{d}$ The observed values, taken from Qi (2000), 
have an estimated uncertainty of 20 \%;
all values refer to the original beam size of the observations
(see Table \ref{tab: nr1})
\label{tab: tab4}
\end{table*}

\subsection{Line profiles and intensities}

The line profiles are calculated using the full 2D radiative
transfer code for the range of depletions derived in \S \ref{sec:
lineratio}. The depletions are further constrained by the absolute
intensities. Specifically, for LkCa~15 $D_C$=5 and 10 with $D_J$=10 is
taken, and for TW~Hya the same $D_J$=10 was assumed but with $D_C$=100
and 200. As a reference, an extra run was performed for LkCa~15 with
no depletions. 
A general turbulent width of 0.2 km s$^{-1}$ is assumed
and the only structured velocity distribution is taken
to be the Keplerian rotation of the disk. This velocity component is
important for the comparison with observations of sources at non-zero
inclination. For these calculations, an inclination of 60\degr \, for
LkCa~15 and 0\degr \, for TW~Hya is used.  The results are convolved
with the appropriate beam as given in Table \ref{tab: nr1}.

A model with no depletion was also run for LkCa 15 with an inclination of 0
\degr \, to check the effect of inclination. Although the total integrated
line intensities changed significantly, their ratios changed only up to 7 \%.
This justifies the approximate radiative transfer approach used in \S 5.2 for
a first estimate of the depletions.

\begin{figure}[h] 
\resizebox{\hsize}{!}{\includegraphics{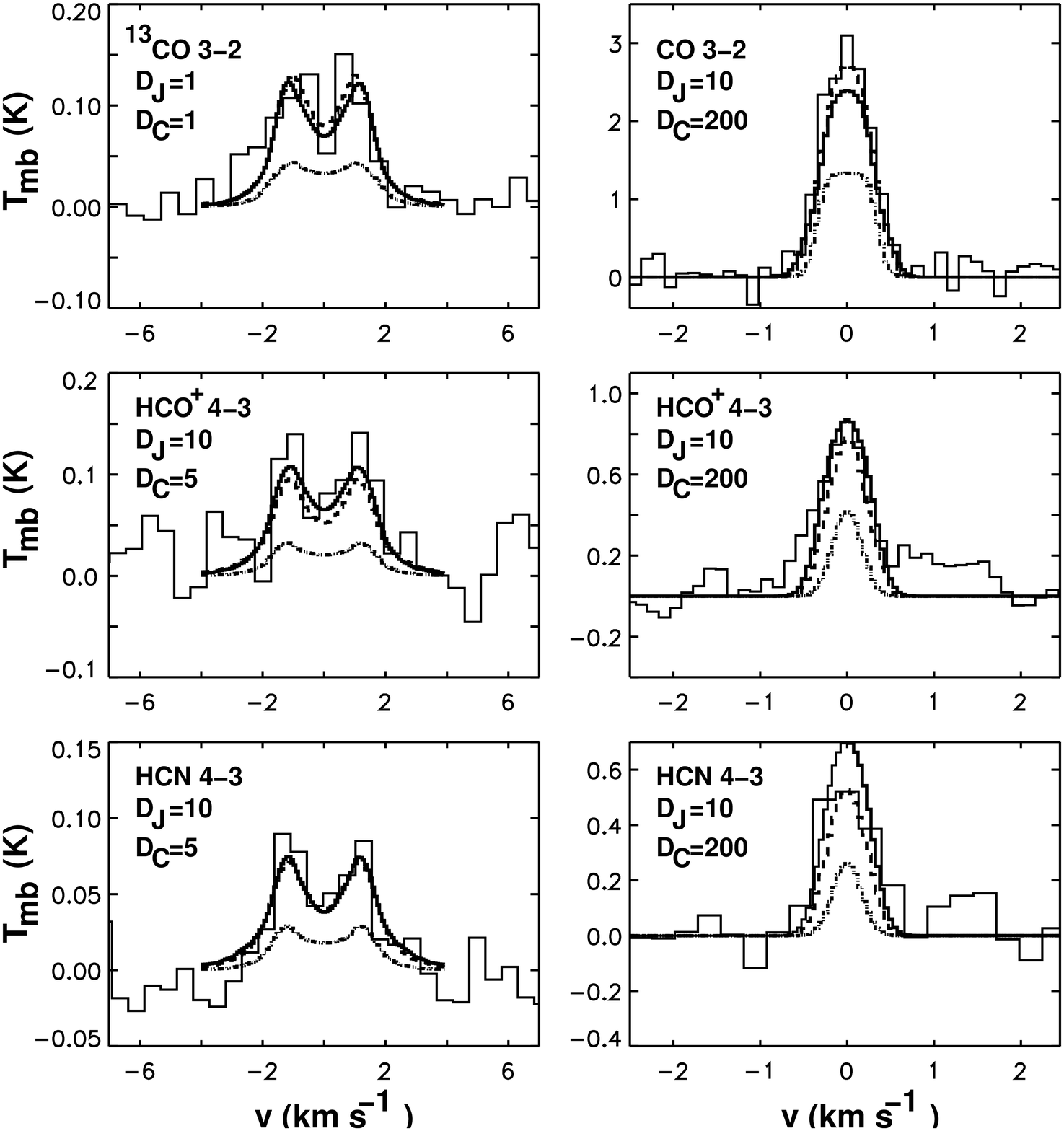}} 
\caption{Observed line profiles for LkCa 15 (left) and TW Hya (right) 
with the three models superposed
(solid: D'Alessio et al.; dashed: Chiang \& Goldreich; dash-dotted: Bell).
The different transitions are indicated with the adopted
depletions compared to standard molecular cloud values. }
\label{fig: colines} 
\end{figure}

The resulting integrated intensities are presented in Table 3 for the
high-$J$ rotational lines and in Table 4 for the lower-$J$
transitions.  For six high-$J$ rotational lines, the observed profiles
are plotted in Figure \ref{fig: colines} with the three calculated
model emission profiles superposed.  In the left-hand figures, three
lines are shown for LkCa~15 whose clear double peaks are due to the
Keplerian rotation in the disk.  On the right, the single peaks for a
face-on disk such as that around TW~Hya are seen. The optically thick
lines from the latter source show that the disk can be fitted with a
turbulent velocity of 0.2 km s$^{-1}$. 

\subsubsection{Depletions}

The absolute intensities in Tables 3 and 4 indicate that refinements
of the inferred depletions are required, since different molecules
favor different amounts of depletion.  Note that the intensities
computed for the cold Bell model are always smaller compared to the
other two models, for both the high and low rotational lines. The
reason for this is twofold. First, for a cold isothermal disk
structure the level populations of any molecule at densities above the
critical density are the same at each position. This means that the
optical depth becomes directly proportional to the column of gas. For
a model with an increasing temperature, the optical depth will not be
directly proportional to the column but smaller. A model with a step
function in its temperature will give results in between these two
cases. Second, the colder disk will have slightly narrower, more
optically thick lines due to the lower thermal motions in the gas,
thereby trapping radiation more effectively.

Both effects are visible in the $^{13}$CO and CO lines (Fig. \ref{fig:
colines}). The two peaks in the $^{13}$CO intensity for the inclined
LkCa 15 disk are reduced significantly compared to the emission at
line center in the Bell model.  In the CO 3--2 line for the face-on TW
Hya disk, the emission predicted by the Bell model is extremely
optically thick, shown by the flat-topped emission profile. Also, the
total linewidth is somewhat smaller compared to the other two disk
models due to the low temperatures.  Thus, the observed line profiles
argue for a rising temperature structure in the vertical direction to
prevent the high optical depths found in the cold isothermal model.

To counteract the low intensities found in the Bell model for the
TW~Hya disk, additional calculations were performed for less severe
depletions (Table 3: $D_J$=10, D$_C=10$).  The integrated intensities
increase to just above the observed values in this case; however, the
lines are extremely optically thick and show nearly square emission
profiles, which is not observed for the CO 3--2 and HCO$^{+}$ 4--3
lines.  Thus, the line profiles speak against small depletions.  In
the two warm disk models, there is no significant difference between
the two assumed depletions and only a slight preference can be given
to $D_C$=100. This is largely based on the HCO$^{+}$/H$^{13}$CO$^{+}$
ratio, which is ill fitted by a depletion of 200 and only moderately
well for $D_C=100$.

For LkCa~15, CO is best fitted with little depletion: all three models
indicate $D_C$ close to 1 for the lower rotational lines.  The upper
limit on the C$^{18}$O 3-2 line and the C$^{18}$O 1-0 emission would
favor some depletion, which, in the case of C$^{18}$O, can be
explained by enhanced photodissociation in the upper layers due to the
lack of self shielding. The CO 3--2 intensity is too low in all three
models, which may be a sign of extended emission beyond 200 AU since
this line is optically thick.  The HCO$^{+}$ and HCN lines are best
fitted by a moderate depletion of $D_C$=5 and $D_J$=10.  For HCN, this
could again be a sign of a lack of shielding against photodissociation
compared to CO. The observed H$^{13}$CN 1--0 is slightly too high for
all three models. Together, the HCN and H$^{13}$CN data indicate that
the HCN abundance needs to be lowered primarily in the surface layer
to both bring the main isotope HCN emission down but keep a high
H$^{13}$CN intensity.

\subsubsection{Probing the temperature and density structure}
\begin{table*}[Ht]
\caption{Line ratios obtained for the three disk models}
\begin{tabular}{lccccccccccccc}
\hline
Line ratio  & Observed & \multicolumn{2}{c}{D'Alessio et al.} 
& \multicolumn{2}{c}{Bell} & \multicolumn{2}{c}{Chiang \& Goldreich} \\
 & LkCa~15$^{c}$  & $D_{C}$, $D_{J}$  & Ratio &$D_{C}$, $D_{J}$  
&Ratio &$D_{C}$, $D_{J}$  & Ratio  & \\
\hline
\vspace{-0.25cm} \\
CO $\frac{6-5}{3-2}$&0.38$^{+ 0.19}_{- 0.13}$ &5,10$^{a}$&0.66 & 1,1$^{a}$
 & 0.36 & 10,10& 0.55 \\  
\vspace{-0.25cm} \\
$^{13}$CO $\frac{3-2}{1-0}$&0.05$^{+ 0.03}_{- 0.02}$            &1,1$^{a}$ &0.06 &10,10$^{a}$& 0.05 & 1,1$^{a}$& 0.05 \\
\vspace{-0.25cm} \\
HCO$^{+}$ $\frac{4-3}{1-0}$& 0.06$^{+ 0.03}_{- 0.02}$           &1,1 & 0.11& 1,1$^{a}$ & 0.06 & 1,1 & 0.08   \\
\vspace{-0.25cm} \\
H$^{13}$CO$^{+}$ $\frac{4-3}{1-0}$& $<$ 1.9
  &10,10$^{b}$&1.44&5,10& 1.00 & 10,10$^{b}$& 1.78&\\
\vspace{-0.25cm} \\
HCN $\frac{4-3}{1-0}$ & 0.08$^{+ 0.04}_{- 0.03}$                &5,10$^{a}$  & 0.09&5,10$^{a}$ & 0.07 &1,1$^{a}$&0.07 \\
\vspace{-0.25cm} \\
\hline 
\vspace{-0.30cm} \\
 & TW Hya$^{c}$ & & &  & \\
\vspace{-0.30cm} \\
\hline
\vspace{-0.25cm} \\
CO $\frac{6-5}{3-2}$& $<$ 1.62  & 100,10 &  0.48 & 100/10,10$^d$ & 0.12/0.27 
& 100,10 &0.41 \\
\vspace{-0.25cm} \\
CO $\frac{4-3}{3-2}$& 2.53$^{+1.26}_{-0.84}$ & 200,10 & 1.28 & 200/10,10$^d$ 
& 1.01/1.13     & 100,10 & 1.29   \\
\vspace{-0.25cm} \\
HCN $\frac{4-3}{3-2}$& 1.09$^{+0.54}_{-0.36}$ & 100,10 & 1.64 & 100/10,10$^d$ 
& 1.38/1.13  & 100,10 & 1.92 \\
\vspace{-0.25cm} \\
\hline

\end{tabular} \\
$^{a}$ Ratios for all three combinations of $D_C$ and $D_J$ 
fall within the error\\
$^{b}$ Ratios for $D_C=5, D_J=10$ and $D_C=10, D_J=10$ fall within the error\\
$^{c}$ The observed values and all ratios refer to the original beam
sizes in which the lines were observed (see Table \ref{tab: nr1}); thus, the
beam size differs between species and lines.
The error bars correspond to a 20 \% uncertainty. \\
$^{d}$ Ratios for high and low values of $D_C$ are shown \\
\label{tab: tab5}
\end{table*}

The calculated line ratios which are sensitive to the temperature and
density distribution are summarized in Table \ref{tab: tab5}.  The
temperature of the upper layer in the LkCa~15 models, as probed by the
CO 6--5/3--2 ratio, fits within the errors to all three models,
confirming the rather cold upper layer of this disk. However, as
explained in the previous section, the absolute intensities are too
low in the cold Bell model. To reproduce the observed CO 6--5
intensity, the CO abundance would have to be increased well above the
cosmic carbon abundance in the cold isothermal model.

For the TW Hya disk, the modeled CO 4--3/3--2 ratios are on the low
side, even in the D'Alessio et al. and Chiang \& Goldreich models,
indicating that the temperature in the surface layers would need to be
higher. However, the calibration uncertainties in the older Kastner et
al.\ (1997) data make this conclusion less firm. Further observations
of high-$J$ CO lines are needed to constrain the temperature structure
of this disk.

The density is probed by the different HCO$^{+}$ and HCN ratios. All
three models are consistent with the observed 4--3/1--0 ratio for the
main isotopes, indicating that the density in the layer above the
midplane is in the correct range. The upper limits on the H$^{13}$CN
and H$^{13}$CO$^+$ prevent any conclusions for the midplane.  As Table
\ref{tab: tab5} shows, the models make different predictions for these
optically thin species and future observations may be used to
distinguish them.

Overall, the absolute line intensities and ratios are consistent with
the models of D'Alessio et al.\ and Chiang \& Goldreich for reasonable
values of the depletions. The current data cannot distinguish between
these two flared disk models.  There is some evidence, however, both
from the line ratios and from the line profiles that the surface of
the disks needs to be warmer than that of a shielded isothermal outer
disk such as computed by Bell (1999).

\section{Discussion}

The high depletions derived for TW~Hya are in agreement with the
conclusions by Kastner et al.\ (1997). The depletion can be due to two
reasons: the first is a change in the gas- to dust-mass ratio to a
value lower than 100 due to removal of gas. The second possibility is
a large depletion of CO and other molecules, but not H$_{2}$. The
former possibility could be partly tested by searches for the pure
rotational H$_2$ lines (Thi et al.\ 2001). Regarding the second
option, the present analysis indicates that the depletion cannot
simply be due to the freezing out of molecules resulting in a large
value of $D_J$: the molecules also need to be depleted in the warmer
upper layers probed by $D_C$.  This latter conclusion could be
consistent with the fact that TW~Hya is a very active UV and X-ray
producing star (Kastner et al.\ 1999), capable of destroying CO due to
dissociation or ionization. This can be tested by measuring the CO
ionization or dissociation products, such as C$^+$, C and CO$^+$.
TW~Hya seems well described by a flared disk model where the
ultraviolet radiation is capable of heating the upper layer (see \S
5.3).

The modeling suggests freezing out with a common value of $D_J$ of at
least 10. The depletion of molecules onto grains can be an important
chemical sink in these disks since most of the mass is cold.  This
makes mass determinations of disks using CO or any of its isotopomers
highly uncertain.

New models have recently been fitted to the SEDs of LkCa~15 and TW~Hya
by Chiang et al.\ (2001) and Chiang (2000).  These models use as one
of the main parameters the dust settling toward the midplane.  For
LkCa~15, the SED modeling indicates that the dust should have settled
within the disk scale height $H$ to explain the observations, with a
high dust temperature in that region ($T_{\rm dust}=$49 K at 100
AU). In these high density regions, the gas and dust temperature
should be coupled; however at heights above the scale height $H$ the
lack of grains will reduce the heating of the gas due to the
photo-electric effect, although some small grains may still be
present.  Together with enhanced cooling due to [O~I] and [C~II], this
 may cause a second temperature inversion with a cool upper layer free
of dust.  The lack of grains in the upper layer would be consistent
with the non-detection of LkCa~15 by HST optical images (K.\
Stapelfeldt, private communication). This suggests a lack of
scattering which should have been readily seen for a flared disk at an
inclination of 60$^{\circ}$ with grains well mixed with the gas and an
albedo of 0.5.  In addition, the models used were calculated for a
stellar temperature of 4000 K, whereas LkCa~15 has a higher effective
temperature (T$_{\mbox{eff}}$=4400 K) which would result in higher
disk dust-temperatures. The relatively low gas temperature in the disk,
indicated by the line observations, strengthen
the conclusions derived from the scattering and SED observations that
dust settling has taken place in LkCa 15.  Self-consistent models of
the gas temperature and abundances of LkCa~15 are needed, taking
dust-settling into account.
 
As noted above, it is not yet possible to distinguish between the
D'Alessio et al. (1999) and the Chiang \& Goldreich (1997) models with
the current observations. The data lack spatial resolution and have
insufficient sensitivity to observe the optically thin isotopic
lines. In addition to higher spatial resolution and sensitivity,
better calibration of the data is needed, all of which will be
provided by the {\it Atacama Large Millimeter Array}.

\section{Conclusions}

The main conclusions from this work are:

\begin{itemize}
\item{High-frequency molecular lines with high critical densities and
excitation temperatures are detected from circumstellar disks.
These observations can be used to test the temperature
and density structure of different disk models in
the literature.}
\item{The $\tau=1$ surfaces of the various lines indicate that the
observed emission of the main isotopes originates from the (warm) intermediate
layer of the disk, whereas the emission from the $^{13}$C isotopes may also
probe the midplane if the molecule is not frozen out completely.}
\item{Most molecules are depleted by a large factor ($>100$) for
TW~Hya and a smaller factor (surface $\approx 1-5$, midplane $\approx
1-50$) for LkCa~15.  Freeze-out onto grains at $T<22$~K is indicated
by the observations, but the molecules are also depleted in the upper,
warmer layers, likely due to photodissociation.  Gaseous species like CO
and its isotopomers should therefore not be used as mass tracers due
to their uncertain abundances.}
\item{ A model with a cold isothermal temperature distribution will
have high optical depths in the lines, thereby reducing the
integrated line emission and producing flat-topped profiles.} 
\item{The TW~Hya disk has a warm surface layer ($>$40~ K) while the
LkCa~15 surface layer is cooler ($\sim$20--40 K). This conclusion depends
sensitively on the calibration accuracy of the high-$J$ CO lines.}
\item{The inferred warm dust from the SED combined with the cooler
gas detected here may be consistent with settling of dust 
in the LkCa~15 disk.}
\item{The density profiles in the three models are consistent with the
observed line ratios, except that the temperature in the upper
layer of a non-irradiated disk such as in the Bell (1999) model is too low
for some sources.}

\end{itemize}

{\it Acknowledgments.} The authors are very grateful to P. D'Alessio,
R. Bell and E. Chiang for sending and discussing the models used in
the paper. They thank M.\ Hogerheijde and F.\ van der Tak for useful
discussions and providing their radiative transfer code, and are
grateful to the staff of the CSO and JCMT for their
support. Astrochemistry in Leiden is supported by a SPINOZA grant from
the Netherlands Organization for Scientific Research (NWO). This
paper is dedicated to Fred Baas, who died on April 4, 2001.  His
expert, generous support at the JCMT was essential to make these
observations possible.


\begin{thebibliography}{99}
\bibitem[Adams, Shu \& Lada 1988]{1988ApJ...326..865A} Adams F. C., 
Shu F. H., \& Lada C. J. 1988, \apj, 326, 865 
\bibitem[Aikawa \& Herbst 1999]{1999A&A...351..233A} Aikawa Y., \&
Herbst E. 1999, \aap, 351, 233
\bibitem[D'Alessio Calvet \& Hartmann 1997]{1997ApJ...474..397D}
D'Alessio P., Calvet N., \& Hartmann L.  1997, \apj, 474, 397
\bibitem[D'Alessio Canto Calvet \& Lizano 1998]{1998ApJ...500..411D} 
D'Alessio P., Cant\'o J., Calvet N., \& Lizano S.  1998, \apj, 500, 411 
\bibitem[D'Alessio et al. 1999a]{1999ApJ...511..896D} D'Alessio P.,
Cant\'o J., Hartmann L., Calvet N., \& Lizano S.  1999,
\apj, 511, 896
\bibitem[D'Alessio et al. 1999b]{1999ApJ...527..893D} D'Alessio P.,
Calvet N., Hartmann L., Lizano S., \& Cant\'o J.  1999,
\apj, 527, 893
\bibitem[Beckwith & Sargent 1996]{1996Natur.383..139B} Beckwith S.\ 
V.\ W., \& Sargent A.\ I.\ 1996, Nature, 383, 139
\bibitem[Beckwith \& Sargent 1993]{1993ApJ...402..280B} Beckwith
S.\ V.\ W., \& Sargent A.\ I.\ 1993, \apj, 402, 280
\bibitem[Beckwith 1999]{Beck2} Beckwith S.\ V.\ W. 1999, The Origin of
Stars and Planetary Systems.  Ed.  C.\ J. Lada, \& N.\ D. Kylafis.
Dordrecht: Kluwer p.579
\bibitem[Bell Cassen Klahr \& Henning 1997]{1997ApJ...486..372B} Bell K. 
R., Cassen P.\ M., Klahr H.\ H., \& Henning T.  1997, \apj, 486, 372 
\bibitem[Bell 1999]{1999ApJ...526..411B} Bell K. R. 1999, \apj, 526, 411 
\bibitem[Bouvier \& Bertout 1992]{1992A&A...263..113B} Bouvier J., \&
Bertout C. 1992, \aap, 263, 113
\bibitem[Burrows et al. 1996]{1996ApJ...473..437B} Burrows C.\ J., et al. 
1996, \apj, 473, 437 
\bibitem[Chiang \& Goldreich 1997]{1997ApJ...490..368C} Chiang E.\ I., \& Goldreich P. 1997, \apj, 490, 368 
\bibitem[Chiang \& Goldreich 1999]{1999ApJ...519..279C} Chiang E. I., \& Goldreich P. 1999, \apj, 519, 279     
\bibitem[Chiang (2000)]{chiang2000} Chiang E.\ I. 2000, Ph.D. thesis,
California Institute of Technology, Pasadena, California
\bibitem[Chiang et al. 2001]{chinag} Chiang E.~I., Joung 
M.~K., Creech-Eakman M.~J., Qi C., Kessler J.~E., Blake G.~A., \& van 
Dishoeck E.~F.\ 2001, \apj, 547, 1077 
\bibitem[Duvert et al. (2000)]{2000A&A...355..165D} Duvert G., 
Guilloteau S., M{\'e}nard F., Simon M., \& Dutrey A. 2000, \aap, 355, 
165 
\bibitem[Dutrey et al. 1996]{dutrey96} Dutrey A., Guilloteau 
S., Duvert G., Prato L., Simon M., Schuster K., \& Menard F.  1996, 
\aap, 309, 493 
\bibitem[Dutrey, Guilloteau \& Guelin (1997)]{1997A&A...317L..55D} Dutrey 
A., Guilloteau S., \& Guelin M. 1997, \aap, 317, L55 
\bibitem[G{\'o}mez \& D'Alessio (2000)]{gomez99} G{\'o}mez 
J., \& D'Alessio P.\ 2000, \apj, 535, 943 
\bibitem[Hartmann \etal (1998)]{hartmann} Hartmann L., Calvet N.,
Gullbring E., \& D'Alessio P. 1998, ApJ, 495, 385
\bibitem[Hogerheijde \& van der Tak \, 2000]{hogerheijde} Hogerheijde
M.\ R., \& van der Tak F. 2000, \aap, 362, 697 
\bibitem[Jansen \etal 1994]{jansen2} Jansen D.\ J., van Dishoeck E.\ F., \& Black J.\ H. 1994, \aap, 282, 605 
\bibitem[Jansen \etal 1996]{jansen3}  Jansen D.\ J., van 
Dishoeck E.\ F., Keene J., Boreiko R.\ T., \& Betz A.\ L.\ 1996, \aap, 
309, 899 
\bibitem[Jansen 1995]{jansen}Jansen D. J. 1995, Ph.D. thesis, Univ. of Leiden
\bibitem[Kastner Zuckerman Weintraub \& Forveille
1997]{1997Sci...277...67K} Kastner J.\ H., Zuckerman B., Weintraub
D.\ A., \& Forveille T. 1997, Science, 277, 67
\bibitem[Kastner et al. (1999)]{1999ApJ...525..837K} Kastner J. H.,
Huenemoerder D.\ P., Schulz N.\ S., \& Weintraub D.\ A. 1999, \apj, 525,
837
\bibitem[Kenyon \& Hartmann (1987)]{1987ApJ...323..714K} Kenyon
S.\ J., \& Hartmann L. 1987, \apj, 323, 714
\bibitem[Krist et al.\ (2000)]{krist2000} Krist J.\ E., 
Stapelfeldt K.\ R., M{\'e}nard F., Padgett D.\ L., \& Burrows C.\ 
J.\ 2000, \apj, 538, 793 
\bibitem[Muzerolle et al.\ 2000]{2000ApJ...535L..47M} Muzerolle J., 
Calvet N., Brice{\~ n}o C., Hartmann L.,  \& Hillenbrand L.\ 2000, \apj, 535, L47
\bibitem[Siess, Forestini, \& Bertout(1999)]{1999A&A...342..480S} Siess 
L., Forestini M., \& Bertout C.\ 1999, \aap, 342, 480 
\bibitem[Stognienko Henning \& Ossenkopf 1995]{1995A&A...296..797S} 
Stognienko R., Henning T., \& Ossenkopf, V. 1995, \aap, 296, 797 
\bibitem[Ossenkopf \& Henning (1994)]{1994A&A...291..943O} Ossenkopf
V., \& Henning T.  1994, \aap, 291, 943
\bibitem[Osterloh \& Beckwith 1995]{1995ApJ...439..288O} Osterloh
M., \& Beckwith S.\ V.\ W. 1995, \apj, 439, 288
\bibitem[Qi et al. (2001)]{Qi2} Qi C. et al., 2001, in preperation
\bibitem[Qi (2000)]{Qi} Qi C. 2000, Ph.D. thesis, California
Institute of Technology, Pasadena, California
\bibitem[Sandford \& Allamandola 1993]{1993ApJ...417..815S} Sandford 
S.\ A., \& Allamandola L.\ J.\ 1993, \apj, 417, 815
\bibitem[]{}
Simon M., Dutrey A., \& Guilloteau S.\ 2000, \apj, 545, 1034
\bibitem[Thi et al. 2001]{wf1} Thi W.\ F. et al., 2001, ApJ, in press
\bibitem[Thamm et al. 1994]{1994A&A...287..493T} Thamm 
E., Steinacker J., \& Henning T.\ 1994, \aap, 287, 493 
\bibitem[Weinberger, et al. (1999)]{weinberger} Weinberger A.\ J., 
Schneider G., Becklin E.\ E., Smith B.\  A., \& Hines D.\  C. 1999, American 
Astronomical Society Meeting, 194, 6904 
\bibitem[Willacy et al. (1998)]{willacy98} 
Willacy K., Klahr H.\ H., Millar T.\ J.\ , \& Henning T.\ 1998, \aap, 
338, 995 
\bibitem[Wilner et al. 2000]{wilner} Wilner D.\ J., Ho P.\ T.\ P., Kastner
J.\ H., \& Rodriguez L.\ F.\ 2000, \apj, 534, L101
\bibitem[Wilson \& Rood (1994)]{1994ARA&A..32..191W} Wilson T.\  L., \&
Rood R. 1994, \araa, 32, 191
\end{thebibliography}
\end{document}